\documentclass[10pt,journal,letterpaper]{IEEEtran}
 
\usepackage{research5}
\usepackage{psfrag}
\usepackage{epsfig}
\usepackage{graphicx}
\usepackage{multicol}
\usepackage{cite} 
\usepackage{amsthm}
\usepackage{subfigure}

\newtheorem{theorem}{Theorem}
\newtheorem{lemma}{Lemma}

\newtheorem{proposition}{Proposition}

\newtheorem{case}{Case}
\newtheorem{scheme}{Scheme}

\newcommand{\e}{\mathop{}\!e}

\newcommand{\ave}{\mathcal{E}}
\renewcommand{\Poisson}{\textnormal{Pois}}

\title{A refined analysis of the Poisson channel in the
  high-photon-efficiency regime}

\author{Ligong Wang,~\IEEEmembership{Member,~IEEE} and Gregory
  W. Wornell,~\IEEEmembership{Fellow,~IEEE}} 

\begin{document}

\maketitle 
\renewcommand{\thefootnote}{} 
\footnotetext[1]{This work was supported in part by the DARPA InPho
  program
 under Contract No. HR0011-10-C-0159, and by AFOSR under Grant
  No.~FA9550-11-1-0183. This paper was presented in part at the 2012 IEEE Information Theory Workshop (ITW) in Lausanne, Switzerland, and will be presented in part at the 2014 IEEE International Symposium on Information Theory (ISIT) in Honolulu, HI, USA.

L. Wang is with the Massachusetts Institute of Technology, Cambridge, MA 02139 USA (e-mail: wlg@mit.edu).

G.W. Wornell is with the Massachusetts Institute of Technology, Cambridge, MA 02139 USA (e-mail: gww@mit.edu).}  \renewcommand{\thefootnote}{\arabic{footnote}}
\begin{abstract}
  We study the discrete-time Poisson channel under the constraint that
  its average input power (in photons per channel use) must not exceed
  some constant $\ave$. We 
  consider the wideband, high-photon-efficiency extreme where $\ave$
  approaches zero, and 
  where the channel's ``dark current'' approaches zero proportionally with
  $\ave$. Improving over a previously obtained first-order
  capacity approximation, we derive a refined approximation,
  which includes the exact characterization of the second-order term, as well as an asymptotic characterization of the third-order term with respect to the dark current. We also show that
  pulse-position modulation is nearly optimal in this regime.
\end{abstract}

\begin{IEEEkeywords}
Optical communication, Poisson channel, channel capacity, wideband, low
SNR.
\end{IEEEkeywords}

\section{Introduction}\label{sec:intro}

\IEEEPARstart{W}{e} consider the discrete-time memoryless Poisson channel whose input
$x$ is in the set $\Reals_0^+$ of nonnegative reals and whose output
$y$ is in the set $\Integers_0^+$ of nonnegative integers. Conditional
on the input $X=x$, the output $Y$ has a Poisson distribution of mean
$(\lambda+x)$, where $\lambda\ge 0$ is called the ``dark
  current'' and is a constant that does not depend on the input~$X$. We
  denote the Poisson distribution of mean $\xi$ by 
$\Poisson_\xi(\cdot)$ so 
\begin{equation}
	\Poisson_\xi(y) = \e^{-\xi} \frac{\xi^y}{y!},\quad y\in\Integers_0^+.
\end{equation}
With this notation the channel law $W(\cdot|\cdot)$ is
\begin{equation}\label{eq:channel_law}
	W(y|x) = \Poisson_{\lambda+x}(y),\quad x\in\Reals_0^+,y\in\Integers_0^+.
\end{equation}

This channel models pulse-amplitude modulated optical communication
where the transmitter sends light signals in \emph{coherent states}
(which are usually produced using laser devices), and where the
receiver employs \emph{direct 
  detection} (i.e., photon counting)~\cite{shapiro09}. The
channel input~$x$ describes the 
expected number 
of \emph{signal photons} (i.e., photons that come from the input light
signal rather than noise) to be detected in the pulse duration, and is
proportional to the light signal's intensity,\footnote{We assume that the pulses are square, which is usually the case in practice. If they are not square, the light signal's intensity should be averaged over the pulse duration.} the pulse duration, the
channel's transmissivity, and the detector's efficiency; see \cite{shapiro09} for details. The channel
output~$y$ is the actual number of photons that 
are detected in the pulse duration. The dark current 
$\lambda$ is the average number of extraneous counts that appear in $y$. We note that, although the name ``dark current'' is traditionally used in the Poisson-channel literature, these extraneous counts usually include both the detector's ``dark clicks'' (which are where the name ``dark current'' comes from) and photons in background radiation.

We impose an \emph{average-power constraint}\footnote{Here ``power''
  is in discrete time, means expected number of detected signal photons per
  channel use, and is proportional to the continuous-time physical
  power times the pulse duration.} on the input  
\begin{equation}\label{eq:ave}
  \E{X} \le \ave
\end{equation}
for some $\ave>0$. 

In applications like free-space
optical communications, the cost of producing and successfully
transmitting photons 
is high, hence high \emph{photon-information efficiency}---amount of information 
transmitted per photon, which we henceforth call simply ``photon efficiency''---is desirable. As we later demonstrate, this
can be achieved in the \emph{wideband} regime, where the
pulse duration of the input approaches 
zero and, assuming that the \emph{continuous-time} average input power is
fixed, where $\ave$ approaches zero proportionally with the
pulse duration. Note that in this regime the average number of
detected background photons or dark clicks also tends to zero
proportionally with the pulse duration. Hence we have the linear
relation 
\begin{equation}\label{eq:lambdac}
  \lambda = c \ave,
\end{equation}
where $c$ is some nonnegative constant that does not change with
$\ave$. In practice, asymptotic results in this regime are useful in scenarios
where $\ave$ is small and where $\lambda$ is comparable to or much
smaller than $\ave$. Scenarios where $\ave$ is small but $\lambda$ is
much larger is better captured by the model where $\lambda$ stays constant
while $\ave$ tends to zero; see
\cite[Proposition 2]{lapidothshapirovenkatesanwang11}. 

We denote the capacity (in nats per channel use) of the channel
\eqref{eq:channel_law} under power constraint \eqref{eq:ave} with
dark current \eqref{eq:lambdac} by $C(\ave, c)$, then
\begin{equation}\label{eq:defC}
  C(\ave,c) = \sup_{\E{X}\le \ave} I(X;Y),
\end{equation}
where the mutual information is computed from the channel law
\eqref{eq:channel_law} and is maximized over input distributions
satisfying~\eqref{eq:ave}, with dark current $\lambda$ given by
\eqref{eq:lambdac}. 
As we shall see, our  
results on the asymptotic behavior of $C(\ave,c)$ hold irrespectively
of whether a \emph{peak-power constraint} 
\begin{equation}\label{eq:peak}
  X \le \mathcal{A} \quad \textnormal{with probability }1
\end{equation}
is imposed or not, as long as $\mathcal{A}$ is positive and does not
approach zero together with $\ave$.

We now formally define the maximum achievable photon efficiency $C_\textnormal{PE}(\ave,c)$, where the subscript notation services a reminder that this quantity is photon efficiency and not capacity:
\begin{equation}
	 C_\textnormal{PE}(\ave,c)  \triangleq 
  \frac{C(\ave,c)}{\ave}. \label{eq:defPE}
\end{equation}

Various capacity results for the discrete-time Poisson channel have
been obtained
\cite{pierce78,massey81,shamai90,lapidothmoser09,lapidothshapirovenkatesanwang11}. Among 
them, \cite[Proposition 1]{lapidothshapirovenkatesanwang11} considers the same
scenario as the present paper and asserts that
\begin{equation}\label{eq:lsvw}
  \lim_{\ave\downarrow 0} \frac{C(\ave,c)}{\ave\log\frac{1}{\ave}}
  = 1,\quad c\in[0,\infty).
\end{equation}
In other words, the maximum photon efficiency satisfies
\begin{equation}
  C_\textnormal{PE}(\ave,c)  = 
  \log\frac{1}{\ave} + o\left(\log\frac{1}{\ave}\right),
  \quad c\in[0,\infty). \label{eq:PE_1st}
\end{equation}
Furthermore, the proof of \cite[Proposition 1]{lapidothshapirovenkatesanwang11} shows that the limit \eqref{eq:lsvw} is achievable using on-off keying. In the Appendix we provide a proof for the converse part of \eqref{eq:lsvw} that is much simpler than the original proof in \cite{lapidothshapirovenkatesanwang11}.\footnote{For the achievability part of \eqref{eq:lsvw}, we note that its proof can be simplified by letting the decoder ignore multiple photons, rather than considering all possible values of $Y$ as in \cite{lapidothshapirovenkatesanwang11}. This can be seen via the achievability proofs in the current paper, which do ignore multiple photons at the receiver, and which yield stronger results than those in \cite{lapidothshapirovenkatesanwang11}.}

The approximation in \eqref{eq:PE_1st} can be compared to the maximum photon
efficiency achievable on the 
\emph{pure-loss bosonic channel} \cite{giovannettiguha04}. This is a quantum optical
communication channel that attenuates the input optical signal, but that does \emph{not} add any noise to it. The transmitter and the receiver of this channel may employ any structure permitted by quantum physics. 
When the transmitter is restricted to sending coherent optical states, and when the receiver is restricted to ideal direct detection (with no dark counts), the pure-loss bosonic channel becomes the Poisson channel \eqref{eq:channel_law} with $\lambda=0$. We denote by
$C_{\textnormal{PE-bosonic}}(\ave)$ the maximum photon efficiency of the
pure-loss bosonic channel under an average-power constraint that is
equivalent to \eqref{eq:ave}. The value of
$C_{\textnormal{PE-bosonic}}(\ave)$ can be easily computed using the
explicit capacity formula \cite[Eq. (4)]{giovannettiguha04}, which
yields 
\begin{equation}\label{eq:PE_bosonic}
  C_{\textnormal{PE-bosonic}}(\ave) = \log\frac{1}{\ave}+1+o(1).
\end{equation}
Comparing \eqref{eq:PE_1st} and \eqref{eq:PE_bosonic} shows the
following:
\begin{itemize}
  \item For the pure-loss bosonic channel in
the wideband regime, coherent-state inputs and direct detection are
optimal 
up to the first-order term in photon efficiency (or, equivalently, in
capacity). For example, they achieve infinite \emph{capacity per
  unit cost} \cite{gallager87,verdu90}.
  \item The dark current does not affect this first-order term. 
\end{itemize}

In the present paper, we refine the analysis in \cite{lapidothshapirovenkatesanwang11} in two aspects. First, we provide a more accurate approximation for $C_\textnormal{PE}(\ave,c)$ that contains higher-order terms, and that reflects the influence of the dark current, i.e., of the constant $c$. Second, we identify near-optimal modulation schemes that facilitate code design for this channel.

Some progress has been made in improving the approximation \eqref{eq:PE_1st}. It is noticed in~\cite{chungguhazheng11} that the photon efficiency
achievable on the Poisson channel~\eqref{eq:channel_law} with $\lambda=0$ may be of the form
\begin{equation}\label{eq:PE_2nd}
  \log\frac{1}{\ave}-\log\log\frac{1}{\ave}+O(1),
\end{equation}
which means that restriction to coherent-state inputs and direct 
detection may induce a loss in the \emph{second-order
  term} in the photon efficiency on the pure-loss bosonic channel.
For practical values of $\ave$, this second-order term can be significant. For example, for $\ave=10^{-5}$, the difference between the first-order term in \eqref{eq:PE_1st} and the first- and
second-order terms in \eqref{eq:PE_2nd} is about
$20\%$. 

The analysis in \cite{chungguhazheng11} (whose main focus is not on the Poisson channel itself) is based on certain assumptions on the input distribution.\footnote{According to conversation with the authors.} It is therefore unclear from \cite{chungguhazheng11} if
\eqref{eq:PE_2nd} is the 
maximum photon efficiency achievable on the Poisson
channel~\eqref{eq:channel_law} with $\lambda=0$ subject to
constraint~\eqref{eq:ave} alone, i.e., if \eqref{eq:PE_2nd} is
the correct expression for $C_{\textnormal{PE}}(\ave,0)$. In the present paper, we prove that this is indeed the case.

Recent works such as \cite{chungguhazheng11,kochmanwornell12} often ignore the dark current in the Poisson channel. It has been unknown to us whether the dark current, i.e., whether the constant $c$ influences the second term in \eqref{eq:PE_2nd} and, if not, whether it influences the next term in photon efficiency. We show that the constant $c$ does \emph{not} influence the second-order term in $C_\textnormal{PE}(\ave,c)$, but \emph{does} influence the third-order, constant term.

It has long been known that that infinite
photon efficiency on the Poisson channel with zero dark current can be
achieved using 
\emph{pulse-position 
  modulation (PPM)} combined with an outer code \cite{pierce78,massey81}. Further, \cite{kochmanwornell12} observes that
PPM can achieve \eqref{eq:PE_2nd} on such a channel. 
PPM greatly simplifies the coding task for this channel, since
one can easily apply existing codes, such as a Reed-Solomon
  code, to the PPM ``super symbols''; while the on-off keying
scheme that achieves 
\eqref{eq:PE_2nd} has a highly skewed input distribution and is hence
difficult to code. The question then arises: how useful is PPM when there is a positive dark current? This question has two parts. First, is PPM still near optimal in terms of capacity (or photon efficiency) when there is a positive dark current? Second, does PPM still simplify coding when there is a dark current? We answer the first part of the question in the affirmative in our main results. We cannot fully answer the second question in this paper, but we shall discuss it in the concluding remarks in Section~\ref{sec:conclusion}.


The rest of this paper is arranged as follows. We introduce some notation and formally define PPM in Section~\ref{sec:notation}. We state and discuss
our main results in Section~\ref{sec:main}. We
then prove the achievability parts of these results in
Sections~\ref{sec:achievability} and~\ref{sec:achievability2}, and prove the converse parts in Section~\ref{sec:converse}. We conclude the paper with numerical comparison of the bounds and some remarks in Section~\ref{sec:conclusion}.

\section{Notation and PPM}\label{sec:notation}
We usually use a lower-case letter like $x$ to denote a constant, and an upper-case letter like $X$ to denote a random variable.

We use natural logarithms, and measure information in nats.

We use the usual $o(\cdot)$ and $O(\cdot)$
notation to describe behaviors of functions
  of $\ave$ in the limit where $\ave$ approaches zero \emph{with other parameters, if any, held fixed}. Specifically,
  given a reference function $f(\cdot)$ (which might be the
  constant~$1$), a function described as $o(f(\ave))$ satisfies
  \begin{equation}
    \lim_{\ave\downarrow 0} \frac{o(f(\ave))}{f(\ave)} = 0,
  \end{equation}
  and a function described as $O(f(\ave))$ satisfies
  \begin{equation}
    \varlimsup_{\ave\downarrow 0}
    \left|\frac{O(f(\ave))}{f(\ave)}\right| < \infty.
  \end{equation}
We emphasize that, in particular, we do \emph{not} use $o(\cdot)$ and $O(\cdot)$ to describe how functions behave with respect to $c$.

We adopt the convention
\begin{equation}
	0 \log 0 = 0.
\end{equation}

We next formally describe what we mean by PPM. On the transmitter side:
\begin{itemize}
  \item The channel uses are divided into frames of equal lengths; 
  \item In each frame, there is only one channel input that is positive, which we call the ``pulse'',  while all the other inputs are zeros; 
  \item The pulses in all frames have the same amplitude.
\end{itemize}
On the receiver side, we distinguish between two cases, which we call \emph{simple PPM} and \emph{soft-decision PPM}, respectively. In simple PPM, the receiver records \emph{at most one pulse} in each frame; if more than one pulse is detected in a frame, then that frame is recorded as an ``erasure'', as if no pulse is detected at all. In soft-decision PPM, the receiver records \emph{up to two pulses} in each frame; frames containing no or more than two detected pulses are treated as erasures. 


\section{Main Results and Discussions}\label{sec:main}
The following theorem provides an approximation for the photon efficiency $C_\textnormal{PE}(\ave,c)$ in the high-photon-efficiency regime.
\begin{theorem}\label{thm:main}
  The maximum photon efficiency $C_{\textnormal{PE}}(\ave,c)$ achievable on the Poisson channel \eqref{eq:channel_law} subject to constraint \eqref{eq:ave} and with dark current \eqref{eq:lambdac}, which we define in \eqref{eq:defPE}, satisfies
  \begin{equation}\label{eq:main}
    C_{\textnormal{PE}}(\ave, c) = \log\frac{1}{\ave} -
    \log\log\frac{1}{\ave} + O(1),\quad c\in[0,\infty).
  \end{equation}
Furthermore, denote
\begin{equation}\label{eq:K}
	K(\ave,c) \triangleq C_\textnormal{PE}(\ave,c) - \log\frac{1}{\ave} + \log\log\frac{1}{\ave},
\end{equation}
then
	\begin{equation}\label{eq:logc}
	\lim_{c\to\infty}\varliminf_{\ave\downarrow 0} \frac{K(\ave,c)}{\log c} = \lim_{c\to\infty} \varlimsup_{\ave\downarrow 0} \frac{K(\ave,c)}{\log c} = -1.
	\end{equation}
\end{theorem}

Our second theorem shows that PPM is nearly optimal in the regime of interest.
\begin{theorem}\label{thm:coding}
The asymptotic expression on the right-hand side of \eqref{eq:main} is achievable with simple PPM. The limits in \eqref{eq:logc} are achievable with soft-decision PPM.
\end{theorem}

The achievability part of Theorem~\ref{thm:main} follows from Theorem~\ref{thm:coding}. To prove the latter, we need to show two things: first, that the largest photon efficiency that is achievable with simple PPM, which we henceforth denote by $C_\textnormal{PE-PPM}(\ave,c)$, satisfies
\begin{equation}\label{eq:lower}
	C_\textnormal{PE-PPM}(\ave,c) \ge \log\frac{1}{\ave} - \log\log\frac{1}{\ave} + O(1),\quad c\in[0,\infty);
\end{equation}
and second, that the maximum photon efficiency that is achievable with soft-decision PPM, which we henceforth denote by $C_\textnormal{PE-PPM(SD)}(\ave,c)$, satisfies
\begin{equation}\label{eq:lower2}
\varliminf_{c\to\infty} \varliminf_{\ave\downarrow 0} \frac{\displaystyle C_\textnormal{PE-PPM(SD)}(\ave,c) - \log\frac{1}{\ave} + \log\log\frac{1}{\ave}}{\log c} \ge -1.
\end{equation}

To prove the converse part of Theorem~\ref{thm:main}, we note that the capacity of the channel is monotonically decreasing in $c$ \cite{lapidothshapirovenkatesanwang11}, and hence so is the photon efficiency. It thus suffices to show two things: first, that, in the absence of dark current, the largest photon efficiency achievable with any scheme satisfies
\begin{equation}\label{eq:upper1}
	C_\textnormal{PE}(\ave,0) \le \log\frac{1}{\ave} -
    \log\log\frac{1}{\ave} + O(1);
\end{equation}
and second, that
\begin{equation}\label{eq:upper}
  \varlimsup_{c\to\infty} \varlimsup_{\ave\downarrow 0} \frac{\displaystyle C_\textnormal{PE}(\ave,c) - \log\frac{1}{\ave} + \log\log\frac{1}{\ave}}{\log c} \le -1.
\end{equation}

We prove \eqref{eq:lower} in
Section~\ref{sec:achievability}, \eqref{eq:lower2} in Section~\ref{sec:achievability2}, and~\eqref{eq:upper1} and \eqref{eq:upper} in
Section~\ref{sec:converse}. 

To better understand the capacity results in Theorem~\ref{thm:main}, we make the following remarks.
\begin{itemize}
  \item Choosing 
    $c=0$ in \eqref{eq:main} confirms that \eqref{eq:PE_2nd} is
    the correct asymptotic expression up to the second-order term for
    $C_{\textnormal{PE}}(\ave,0)$. Compared to~\eqref{eq:PE_bosonic}
    this means that, on the pure-loss bosonic channel, for small $\ave$, restricting 
    the receiver to using direct detection induces a loss in photon
    efficiency of about $\log\log\frac{1}{\ave}$ nats per photon. Note
    that 
    the capacity of the pure-loss bosonic channel can be achieved
    using coherent input states only \cite{giovannettiguha04}, so this
    loss is indeed due to 
    direct detection, and not due to coherent input states. 

Attempts
    to overcome this loss by employing other feasible detection
    techniques have so far been unsuccessful 
    \cite{chungguhazheng11,dolinar11}. 
  \item The expression on the right-hand side 
    of \eqref{eq:main} does not depend on $c$, so the value
    of $c$ affects neither the first-order term nor the second-order
    term in 
    $C_{\textnormal{PE}}(\ave,c)$. In particular, these two terms do
    not depend on
    whether $c$ is zero or positive. 

\item The first term in $C_\textnormal{PE}(\ave,c)$ that $c$ does affect is the third, constant term. Indeed, though we have not given an exact expression for the constant term, the asymptotic property \eqref{eq:logc} shows that $c$ affects the constant term in such a way that, for large $c$, the constant term is approximately $-\log c$.
\item Theorem~\ref{thm:main} suggests the approximation
\begin{equation}\label{eq:approx}
	C_\textnormal{PE}(\ave,c) \approx \log\frac{1}{\ave} - \log\log\frac{1}{\ave} - \log c.
\end{equation}
The terms constituting the approximation error, roughly speaking, are either vanishing for small $\ave$, or small compared to $\log c$ for large $c$. Note that if we fix the dark current, i.e., if we fix the product $c\ave$,  then the first and third terms on the right-hand side of \eqref{eq:approx} cancel. This is intuitively in agreement with \cite[Proposition 2]{lapidothshapirovenkatesanwang11}, which asserts that, for fixed dark current, photon efficiency scales like some constant times $\log\log (1/\ave)$ and hence \emph{not} like $\log(1/\ave)$. We note, however, that \eqref{eq:logc} cannot be derived directly using \eqref{eq:main} together with \cite[Proposition 2]{lapidothshapirovenkatesanwang11}, because we cannot change the order of the  limits. In \eqref{eq:logc} we do not let $\ave$ tend to zero and $c$ tend to infinity simultaneously. Instead, we first let $\ave$ tend to zero to close down onto the constant term in $C_\textnormal{PE}(\ave,c)$ with respect to $\ave$, and then let $c$ tend to infinity to study the asymptotic behavior of this constant term with respect to $c$.
\item The approximation \eqref{eq:approx} is good for large $c$, but diverges as $c$ tends to zero. We hence need a better approximation for the constant term, which behaves like $-\log c$ for large $c$, but which does not diverge for small $c$. As both the nonasymptotic bounds and the numerical simulations we show later will suggest, $-\log(1+c)$ is a good approximation:
\begin{equation}\label{eq:log1c}
	C_\textnormal{PE}(\ave,c) \approx \log\frac{1}{\ave} - \log\log\frac{1}{\ave} - \log (1+c).
\end{equation}
\end{itemize}

For modulation and coding considerations, we make some remarks following Theorem~\ref{thm:coding}.
\begin{itemize}
  \item Theorem~\ref{thm:coding} shows that PPM is optimal up to the second-order term in photon efficiency when $c=0$. Hence, compared to the restriction to the receiver using direct detection, the further restriction to PPM induces only a small additional loss in photon efficiency.

  \item Furthermore, for the third-order, constant term, (soft-decision) PPM is also not far from optimal, in the sense that it achieves the optimal asymptotic behavior of this term for large~$c$.  
\item  In Section~\ref{sec:achievability} we show that
    \begin{IEEEeqnarray}{rCl}
      \lefteqn{C_{\textnormal{PE-PPM}}(\ave,c)} ~\nonumber\\
	& \ge & \log\frac{1}{\ave} -
      \log\log\frac{1}{\ave} - c -\log(1+c)- \frac{3}{2} + o(1).\IEEEeqnarraynumspace \label{eq:PPM_strong}
    \end{IEEEeqnarray}
    A careful analysis will confirm that the bound
    \eqref{eq:PPM_strong} is tight in the regime of interest, in the sense that simple PPM cannot achieve a constant term that is better than linear in $c$ (while being second-order optimal). This is in contrast to soft-decision PPM, which can achieve a constant term that is logarithmic in $c$. In particular, simple PPM is clearly \emph{not} third-order optimal.\footnote{A scheme between simple and soft-decision PPM is the following. When detecting multiple pulses in a frame, the receiver randomly selects and records one of the positions (possibly together with a ``quality'' flag). This scheme outperforms simple PPM in photon efficiency, but its third-order term is still linear in $c$: it scales like $-c/2$ instead of $-c$.}
  \item In the PPM schemes that achieve \eqref{eq:lower} and
    \eqref{eq:lower2}, which we describe in
    Sections~\ref{sec:achievability} and~\ref{sec:achievability2}, the pulse has amplitude
    $1/\bigl(\log(1/\ave)\bigr)$, which depends on $\ave$, and which
    tends to zero as $\ave$ tends to zero. An on-off keying scheme having the same amplitude for its ``on'' signals can also achieve these lower bounds, because the dependence between the channel inputs introduced by PPM can only reduce the total mutual information between channel inputs and outputs. These (PPM and on-off keying schemes) are different from
    the on-off keying scheme used in
    \cite{lapidothshapirovenkatesanwang11} where the ``on'' signal has
    a fixed amplitude that does not depend on $\ave$. The latter on-off
    keying scheme, as well as any PPM scheme with a fixed pulse
    amplitude (with respect to $\ave$), is not second-order optimal on the Poisson channel.  
  \item Because in the PPM schemes to achieve \eqref{eq:lower} and \eqref{eq:lower2} the pulse
    tends to zero as $\ave$ tends to zero, we know that, as claimed in
    Section~\ref{sec:intro}, both our theorems hold if a constant (i.e.,
    not approaching zero together with $\ave$) peak-power constraint
    as in \eqref{eq:peak} is imposed on $X$ in addition to \eqref{eq:ave}.

\end{itemize}

We next proceed to prove our main results, and leave further discussions to Section~\ref{sec:conclusion}.

\section{Proof of the Simple-PPM Lower Bound
  \eqref{eq:lower}}\label{sec:achievability} 

Assume that $\ave$ is small enough so that
\begin{subequations}\label{eq:lower1_conditions}
\begin{IEEEeqnarray}{rCl}
	\ave\log\frac{1}{\ave} & < & 1\\
	\ave & < & \e^{-c}.
\end{IEEEeqnarray}
\end{subequations}
Consider the
following simple PPM scheme:
\begin{scheme}\label{sch:PPM}
\mbox{}
\begin{itemize}

  \item The channel uses are divided into frames of length $b$, so each frame
    contains $b$ input symbols $x_1,\ldots,x_b$ and $b$ corresponding
    output symbols $y_1,\ldots,y_b$. We set 
    \begin{equation}\label{eq:b}
      b = \left\lfloor \frac{1}{\ave\log\frac{1}{\ave}} \right\rfloor.
    \end{equation}
  \item Within each length-$b$ frame, there is always one input that
    equals $\eta$, and all the other $(b-1)$ inputs are zeros. 
    Each frame is then fully specified by the position of its
    unique nonzero symbol, i.e., its pulse position. We
    consider each frame as a 
    ``super input symbol'' $\tilde{x}$ that takes value in
    $\{1,\ldots,b\}$. Here $\tilde{x}=i$, $i\in\{1,\ldots,b\}$, means
    \begin{subequations}
    \begin{IEEEeqnarray}{rCl}
      x_i & = & \eta\\
      x_j & = & 0,\quad j\in\{1,\ldots,b\}, j\neq i.
    \end{IEEEeqnarray}
    \end{subequations}
    To meet the
    average-power constraint \eqref{eq:ave} with equality, we
    require
    \begin{equation}\label{eq:beta}
      \eta = b\ave.
    \end{equation}
  \item The $b$ output
    symbols $y_1,\ldots,y_b$ are mapped to one ``super output symbol''
    $\tilde{y}$ that takes 
    value in $\{1,\ldots,b, {?}\}$ in the
    following way: $\tilde{y}=i$, $i\in\{1,\ldots,b\}$, if $y_i$ is
    the \emph{unique} nonzero term in $\{y_1,\ldots,y_b\}$; and
    $\tilde{y}={?}$ if there is more than one or no nonzero term in
    $\{y_1,\ldots,y_b\}$. 
\end{itemize}
\end{scheme}

We have the following lower bound on the photon efficiency achieved by the above scheme.

\begin{proposition}\label{prp:1}
Under the conditions \eqref{eq:lower1_conditions}, Scheme~\ref{sch:PPM} achieves photon efficiency
\begin{IEEEeqnarray}{rCl}
  	\lefteqn{C_{\textnormal{PE-PPM}}(\ave,c)}~~~~~\nonumber\\ & \ge & \left(1-\frac{\eta}{2}\right) \log b - c\eta \log b - \left(1+\frac{c\ave}{\eta}\right) \log (1+c)\nonumber\\
	& & {} + \left(1+\frac{c\ave}{\eta}\right) \left\{\log(1-c\eta) + \log\left(1-\frac{\eta}{2}\right)\right\} \nonumber\\ & & {}- 1 - \frac{c\ave}{\eta}\label{eq:prp1}
\end{IEEEeqnarray}
where $b$ and $\eta$ are given in \eqref{eq:b} and \eqref{eq:beta}, respectively.
\end{proposition}

\begin{IEEEproof}[Proof of \eqref{eq:lower} using Proposition~\ref{prp:1}]
We plug \eqref{eq:b} and \eqref{eq:beta} into \eqref{eq:prp1}, note that
\begin{equation}
	\left\lfloor \frac{1}{\ave\log\frac{1}{\ave}}\right\rfloor = \frac{1}{\ave\log\frac{1}{\ave}} + O(1)
\end{equation}
and simplify \eqref{eq:prp1} to obtain
\begin{IEEEeqnarray}{rCl}
	\lefteqn{C_\textnormal{PE-PPM}(\ave,c)}~~~\nonumber\\ & \ge & \log\frac{1}{\ave} - \log\log\frac{1}{\ave} - c -\log (1+c) - \frac{3}{2} + o(1)\IEEEeqnarraynumspace\\
	& = & \log\frac{1}{\ave} - \log\log\frac{1}{\ave} +O(1).
\end{IEEEeqnarray}
\end{IEEEproof}

When proving Proposition~\ref{prp:1}, as well as Propositions~\ref{prp:2} and~\ref{prp:upper}, we frequently use the inequalities summarized in the following lemma.
\begin{lemma}
For all $a\ge 0$,
\begin{IEEEeqnarray}{rCl}
	\log(1+a) & \le & a, \label{eq:log}\\
	1 - e^{-a} & \le & a,\label{eq:exp}\\
	1-e^{-a} & \ge & a - \frac{1}{2}a^2.\label{eq:exp2}
\end{IEEEeqnarray}
\end{lemma}

\begin{IEEEproof}[Proof of Proposition \ref{prp:1}]
We compute the transition matrix of the PPM ``super
channel'' as follows:
\begin{IEEEeqnarray}{rCl}
  \tilde{W}(i|i) & = &
  \Pr[\tilde{Y}=i |\tilde{X}=i]\\
  & = & \Pr\left[ \left.Y_i\ge 1\right|X_i = \eta \right] \prod_{\substack{k=1\\k\neq i}}^b \Pr
  \left[ \left. Y_k=0\right.|X_k=0 \right] \IEEEeqnarraynumspace \\
  & = & \bigl(1-W(0|\eta)\bigr) \bigl(W(0|0)\bigr)^{b-1}\\
  & = & (1-\e^{-\eta-c\ave})\e^{-(b-1)c\ave}\\
  & = & \e^{-(b-1)c\ave} - \e^{-\eta-bc\ave} \\
  & \triangleq & p_0, \qquad i \in
  \{1,\ldots,b\}; \label{eq:defp0}\\[3mm]
  \tilde{W}(j|i) & = &
  \Pr[\tilde{Y}=j |\tilde{X}=i ]\\ 
  & = & \Pr\left[ \left.Y_i=0 \right|X_i = \eta \right] \Pr\left[
    \left.Y_j\ge 1\right|X_j = 0 \right]\nonumber \\
  && \cdot \prod_{\substack{k=1\\k\notin\{i,j\}}}^b
  \Pr\left[ \left.Y_k=0\right|X_k = 0 \right]\\
  & = & W(0|\eta)\bigl(1-W(0|0)\bigr)\bigl(W(0|0)\bigr)^{b-2}\\
  & = & \e^{-\eta-c\ave} (1-\e^{-c\ave}) \e^{-(b-2)c\ave}\\
  & = & \e^{-\eta-(b-1)c\ave} - \e^{-\eta - bc\ave}\\
  & \triangleq & p_1, \qquad i,j\in\{1,\ldots,b\},\ i\neq j; \label{eq:defp1} \\[3mm]
  \tilde{W}({?}|i) & = & 1-p_0-(b-1)p_1,\qquad i
  \in\{1, \ldots, b\}. 
\end{IEEEeqnarray}
We now have, irrespective of the distribution of $\tilde{X}$, 
\begin{IEEEeqnarray}{rCl}
	H(\tilde{Y}|\tilde{X}) & = & p_0 \log\frac{1}{p_0} + (b-1)p_1 \log\frac{1}{p_1} \nonumber\\
	& & {} + \bigl(1-p_0-(b-1)p_1\bigr) \log\frac{1}{1-p_0-(b-1)p_1}.\nonumber\\
	& & \,
\end{IEEEeqnarray}
Denote the capacity of this super channel by
$\tilde{C}(\ave,c,b,\eta)$, then
\begin{equation}\label{eq:Ctild}
  \tilde{C}(\ave,c,b,\eta) = \max_{P_{\tilde{X}}}
  I(\tilde{X};\tilde{Y}).
\end{equation}
Note that the total input power (i.e., expected number of detected
signal photons) in each frame equals $\eta$. Therefore we have the following
lower bound on 
$C_{\textnormal{PE-PPM}}(\ave,c)$: 
\begin{equation}\label{eq:super}
  C_{\textnormal{PE-PPM}} (\ave, c) \ge
  \frac{\tilde{C}(\ave,c,b,\eta)}{\eta}.
\end{equation}
It can be easily verified that the optimal input distribution
for~\eqref{eq:Ctild} is the uniform distribution 
\begin{equation}
  P_{\tilde{X}}(i) = \frac{1}{b},\quad i \in\{1,\ldots,b\},
\end{equation}
which induces the following marginal distribution on $\tilde{Y}$:
\begin{subequations}
\begin{IEEEeqnarray}{rCl}
  P_{\tilde{Y}}(i) & = &  \frac{p_0+(b-1)p_1}{b}, \quad
  i\in\{1,\ldots,b\},\\
  P_{\tilde{Y}}({?}) & = & 1-p_0-(b-1)p_1.
\end{IEEEeqnarray}
\end{subequations}
We can now use the above joint distribution on
$(\tilde{X},\tilde{Y})$ to lower-bound
$\tilde{C}(\ave,c,b,\eta)$ as follows:
\begin{IEEEeqnarray}{rCl}
  \lefteqn{\tilde{C}(\ave,c,b,\eta)}~~~~\nonumber\\
  & = & I(\tilde{X};\tilde{Y})\\
  & = & H(\tilde{Y})-H( 
    \tilde{Y}|\tilde{X} )\\
  & = & (1-p_0-(b-1)p_1)\log\frac{1}{1-p_0-(b-1)p_1} \nonumber\\
  && {} +
  (p_0+(b-1)p_1)\log\frac{b}{p_0+(b-1)p_1}\nonumber \\
  & & {}-(1-p_0-(b-1)p_1)\log\frac{1}{1-p_0-(b-1)p_1} \nonumber\\
  && {} -
  p_0\log\frac{1}{p_0} - (b-1) p_1\log\frac{1}{p_1}\\
  & = & p_0 \log\frac{b p_0}{p_0+(b-1)p_1} + (b-1) p_1\log
  \frac{bp_1}{p_0+(b-1)p_1}\nonumber\\ & & \, \label{eq:PPM_exact} \\
  & = & p_0\log b - p_0 \underbrace{\log
  \frac{p_0+(b-1)p_1}{p_0}}_{\le \log \left(1+\frac{b p_1}{p_0}\right)} \nonumber\\
  & & {} - (b-1) p_1 \underbrace{\log
  \frac{p_0+(b-1)p_1}{bp_1}}_{ = \log\left(1+\frac{p_0-p_1}{bp_1}\right)\le \frac{p_0-p_1}{bp_1}}\\
  & \ge & p_0\log b - p_0\log \left(1+\frac{b p_1}{p_0}\right) -
  \underbrace{\frac{b-1}{b}}_{\le 1} \underbrace{(p_0-p_1)}_{\le p_0}
  \IEEEeqnarraynumspace \\ 
  & \ge & p_0 \log b - p_0\log \left(1+\frac{b p_1}{p_0}\right) - p_0.\label{eq:Ctilde}
\end{IEEEeqnarray}
Next note that $p_0$ can be upper-bounded as
\begin{IEEEeqnarray}{rCl}
  p_0 & = & \e^{-(b-1)c\ave} - \e^{-\eta-bc\ave}\\
  & = &
  \e^{-(b-1)c\ave}
  \left(1-\e^{-\eta-c\ave}\right) \\
  & \le & 1-\e^{-\eta-c\ave}\\
  & \le & \eta + c \ave, \label{eq:p0}
\end{IEEEeqnarray}
where the last step follows from \eqref{eq:exp}. It can also be lower-bounded as
\begin{IEEEeqnarray}{rCl}
  p_0 & = & \e^{-(b-1)c\ave} - \e^{-\eta-bc\ave}\\
  & \ge & \e^{-bc\ave} - \e^{-\eta-bc\ave}\\
  & = & \underbrace{\e^{-bc\ave}}_{\ge 1-bc\ave}
  \underbrace{(1-\e^{-\eta})}_{\ge
      \eta-\frac{1}{2}\eta^2} \\
  & \ge & (1-bc\ave)\left(\eta-\frac{1}{2}\eta^2\right),\label{eq:p02}
\end{IEEEeqnarray}
where the last inequality follows from \eqref{eq:exp} and \eqref{eq:exp2}, and from the fact that the first multiplicand on its right-hand side is (in fact, both multiplicands are)
nonnegative for $\ave$ satisfying \eqref{eq:lower1_conditions} and other parameters chosen accordingly.
Also note that $p_1$ can be upper-bounded as
\begin{IEEEeqnarray}{rCl}
  p_1 & = & \e^{-\eta-(b-1)c\ave} - \e^{-\eta - bc\ave}\\ 
  & = & \underbrace{\e^{-\eta-(b-1)c\ave}}_{\le 1}
  \underbrace{\left(1-\e^{-c\ave}\right)}_{\le c\ave}\\
  & \le & c\ave.\label{eq:p1}
\end{IEEEeqnarray}
Using \eqref{eq:p0}, \eqref{eq:p02} and \eqref{eq:p1} we can continue
the chain of inequalities \eqref{eq:Ctilde} to further lower-bound
$\tilde{C}(\ave,c,b,\eta)$ as
\begin{IEEEeqnarray}{rCl}
  \lefteqn{\tilde{C}(\ave,c,b,\eta)}\nonumber\\
  & \ge & (1-bc\ave)\left(\eta-\frac{1}{2}\eta^2\right)\log b \nonumber\\
	& & {} -(\eta+c\ave) \log\left(1+\frac{bc\ave}{(1-bc\ave)\left(\eta-\frac{1}{2}\eta^2\right)}\right) \nonumber\\
	& & {} - \eta -c\ave \IEEEeqnarraynumspace \\ 
  & \ge & (1-bc\ave)\left(\eta-\frac{1}{2}\eta^2\right)\log b - (\eta+c\ave)\log\left(1+\frac{bc\ave}{\eta} \right)\nonumber\\
	& & {} + (\eta+c\ave) \left\{\log(1-bc\ave) + \log\left(1-\frac{\eta}{2}\right)\right\} \nonumber\\
	& & {} - \eta - c\ave \label{eq:Ctilde_2ndlast}\\
  & \ge & \left(\eta-\frac{1}{2}\eta^2\right)\log b -bc\ave\eta \log b - (\eta+c\ave)\log\left(1+\frac{bc\ave}{\eta}\right) \nonumber\\
	& & {} + (\eta+c\ave) \left\{\log(1-bc\ave) + \log\left(1-\frac{\eta}{2}\right)\right\} \nonumber\\
	& & {} - \eta - c\ave. \label{eq:Ctilde_last} 
\end{IEEEeqnarray}
Here, \eqref{eq:Ctilde_2ndlast} follows from the fact that, for all $\alpha\ge 0$ and $\beta\ge 1$,
\begin{equation}
	\log (1+\alpha\beta) \le \log(\beta+\alpha\beta) =\log \beta + \log(1+\alpha),
\end{equation}
with the choices
\begin{IEEEeqnarray}{rCl}
	\alpha & = & \frac{bc\ave}{\eta},\\
	\beta & = & \frac{1}{(1-bc\ave)\left(1-\frac{1}{2}\eta\right)},
\end{IEEEeqnarray}
which indeed satisfy $\alpha\ge 0$ and $\beta\ge 1$; and \eqref{eq:Ctilde_last} follows by dropping the term $\frac{1}{2}bc\ave\eta^2\log b$, which is positive.

Combining \eqref{eq:beta}, \eqref{eq:super}, and \eqref{eq:Ctilde_last} yields
 \eqref{eq:prp1}. 
\end{IEEEproof}


\section{Proof of the Soft-Decision-PPM Lower Bound~\eqref{eq:lower2}}\label{sec:achievability2}

Consider the following soft-decision PPM scheme:

\begin{scheme}\label{sch:PPM2}
	The transmitter performs the same PPM as in Scheme~\ref{sch:PPM}. The receiver maps the $b$ output symbols to a ``super symbol'' $\hat{y}$ that takes value in $\{1,\ldots, b, {?}\}\cup \bigl\{\{i,j\}\subset\{1,\ldots,b\}\bigr\}$. The mapping rule is as follows. Take $\hat{y}=i$ if $y_i$ is the \emph{unique} nonzero term in $\{y_1,\ldots,y_b\}$; take $\hat{y}=\{i,j\}$ if $y_i$ and $y_j$ are the \emph{only two} nonzero terms in $\{y_1,\ldots,y_b\}$; if there are more than two or no nonzero term in $\{y_1,\ldots,y_b\}$, take $\hat{y}={?}$.
\end{scheme}

\begin{proposition}\label{prp:2}
Under the conditions \eqref{eq:lower1_conditions}, Scheme~\ref{sch:PPM2} achieves photon efficiency
\begin{IEEEeqnarray}{rCl}
  	\lefteqn{C_{\textnormal{PE-PPM(SD)}}(\ave,c)}~~\nonumber\\ & \ge & \left(1-\frac{\eta}{2}\right) \log b - c\eta \log b - \left(1+\frac{c\ave}{\eta}\right) \log (1+c)\nonumber\\
	& & {} + \left(1+\frac{c\ave}{\eta}\right) \left\{\log(1-c\eta) + \log\left(1-\frac{\eta}{2}\right)\right\} - 1 - \frac{c\ave}{\eta} \nonumber\\
	& & {} + \left(1-\frac{\eta}{2}\right)(b-1) \left(c\ave - \frac{c^2 \ave^2}{2}\right) (1-c\eta) \log\frac{b}{2}\nonumber\\
& & {} - \frac{c^2}{2}\eta - c(\eta+c\ave) \label{eq:prp2}
\end{IEEEeqnarray}
where $b$ and $\eta$ are given in \eqref{eq:b} and \eqref{eq:beta}, respectively.
\end{proposition}

\begin{IEEEproof}[Proof of \eqref{eq:lower2} using Proposition \ref{prp:2}]
Simplifying \eqref{eq:prp2} we obtain
\begin{IEEEeqnarray}{rCl}
\lefteqn{\varliminf_{\ave\downarrow 0} \left\{C_\textnormal{PE-PPM(SD)}(\ave,c) - \log\frac{1}{\ave} + \log\log\frac{1}{\ave}\right\}}~~~~~~~~~~~~~~~~~~~~~~~~~~~~~~~\nonumber\\
	& \ge & -\log(1+c) - \frac{3}{2}.
\end{IEEEeqnarray}
This immediately yields \eqref{eq:lower2}.
\end{IEEEproof}

\begin{IEEEproof}[Proof of Proposition~\ref{prp:2}]
We compute the transition matrix of the super channel that results from Scheme~\ref{sch:PPM2}. We first note
\begin{IEEEeqnarray}{rCl}
	\hat{W}(i|i) & = & p_0, \\
	\hat{W}(j|i) & = & p_1, \quad i,j\in\{1,\ldots,b\}, \ i\neq j,
\end{IEEEeqnarray}
where $p_0$ and $p_1$ are given in \eqref{eq:defp0} and \eqref{eq:defp1}, respectively. For the remaining elements of the transition matrix we have
\begin{IEEEeqnarray}{rCl}
	\hat{W}\bigl(\bigl. \{i,j\}\bigr| i\bigr) & = & \Pr[Y_i\ge1|X_i=\eta] \Pr [Y_j\ge1|X_j=0]\nonumber\\
	& & {} \cdot \prod_{\substack{\ell=1\\ \ell\notin \{i,j\}}}^b \Pr[Y_\ell=0|X_\ell=0]\\
	& = & (1-\e^{-\eta-c\ave}) (1-\e^{-c\ave}) \e^{-(b-2)c\ave}\\
	& \triangleq & p_2,\quad \{i,j\}\subset\{1,\ldots,b\};\label{eq:defp2}\\[300mm]
	\hat{W}\bigl(\bigl. \{j,k\}\bigr| i\bigr) & = & \Pr[Y_i=0|X_i=\eta] \Pr [Y_j\ge1|X_j=0] \nonumber\\
	& & \cdot\Pr[Y_k\ge1|X_k=0] \nonumber\\ & & \cdot \prod_{\substack{\ell=1\\\ell\notin\{i,j,k\}}}^b \Pr[Y_\ell=0|X_\ell=0]\\
	& = & \e^{-\eta-c\ave} (1-\e^{-c\ave})^2 \e^{-(b-3)c\ave}\\
	& \triangleq & p_3,\quad \{i,j,k\}\subset \{1,\ldots,b\}; \label{eq:defp3}\\
	\hat{W}({?}|i) & = & 1-p_0-(b-1)p_1-(b-1)p_2\nonumber\\
	& & {} -\frac{(b-1)(b-2)}{2}p_3\\
	& \triangleq & p_4,\quad i \in\{1,\ldots,b\}. \label{eq:defp4}
\end{IEEEeqnarray}
We now have, irrespective of the distribution of $\tilde{X}$,
\begin{IEEEeqnarray}{rCl}
	H(\hat{Y}|\tilde{X}) & = & p_0\log\frac{1}{p_0} + (b-1)p_1\log\frac{1}{p_1}\nonumber\\
	&&{}+ (b-1)p_2\log\frac{1}{p_2} + \frac{(b-1)(b-2)}{2}p_3\log\frac{1}{p_3} \nonumber\\
	& & {} + p_4\log\frac{1}{p_4}.\label{eq:HYhatX}
\end{IEEEeqnarray}

Choosing a uniform $\tilde{X}$ (which is optimal as in Section~\ref{sec:achievability}), we have the following distribution on $\hat{Y}$
\begin{IEEEeqnarray}{rCl}
	P_{\hat{Y}} (i )& = & \frac{p_0+(b-1)p_1}{b},\quad i\in\{1,\ldots,b\};\\
	P_{\hat{Y}}\bigl(\{i,j\}\bigr) & = & \frac{2p_2 + (b-2)p_3}{b}, \quad \{i,j\}\subset\{1,\ldots,b\};\IEEEeqnarraynumspace\\
	P_{\hat{Y}}({?}) & = & p_4.
\end{IEEEeqnarray}
Therefore
\begin{IEEEeqnarray}{rCl}
	\lefteqn{H(\hat{Y})}~\nonumber\\ & = & \bigl(p_0+(b-1)p_1\bigr) \log\frac{b}{p_0+(b-1)p_1} \nonumber\\
	& & {} + \left((b-1)p_2 + \frac{(b-1)(b-2)}{2}p_3\right) \log \frac{b}{2p_2 + (b-2)p_3} \nonumber\\
	& & {} + p_4\log\frac{1}{p_4}. \label{eq:HYhat}
\end{IEEEeqnarray}

Using \eqref{eq:HYhatX} and \eqref{eq:HYhat} we have
\begin{IEEEeqnarray}{rCl}
	I(\tilde{X};\hat{Y}) & = & H(\hat{Y}) - H(\hat{Y}|\tilde{X})\\
	& = & p_0\log\frac{bp_0}{p_0+(b-1)p_1}\nonumber\\
	& & {} + (b-1)p_1\log\frac{bp_1}{p_0+(b-1)p_1}\nonumber\\
	& & {} + (b-1)p_2 \log\frac{bp_2}{2p_2+(b-2)p_3}\nonumber\\
	& & {} + \frac{(b-1)(b-2)}{2}p_3 \log\frac{bp_3}{2p_2+(b-2)p_3}.\IEEEeqnarraynumspace\label{eq:IYhat}
\end{IEEEeqnarray}
At this point, note that the first two summands on the right-hand side of \eqref{eq:IYhat} constitute $I(\tilde{X};\tilde{Y})$, which we analyzed in Section~\ref{sec:achievability}. It remains to find lower bounds on the last two summands. We lower-bound the third term on the right-hand side of \eqref{eq:IYhat} as
\begin{IEEEeqnarray}{rCl}
\lefteqn{(b-1)p_2 \log\frac{bp_2}{2p_2+(b-2)p_3}}~~\nonumber\\
	& = & (b-1)p_2\log \frac{b}{2} - (b-1)p_2\underbrace{\log\frac{p_2 + \frac{b-2}{2}p_3}{p_2}}_{=\log\left(1+\frac{(b-2)p_3}{2p_2}\right)\le \frac{(b-2)p_3}{2p_2}} \IEEEeqnarraynumspace\\
	& \ge & (b-1)p_2\log\frac{b}{2} - \frac{(b-1)(b-2)}{2}p_3\\
	& \ge & (b-1)p_2\log\frac{b}{2} - \frac{b^2}{2}p_3.\label{eq:PPM2third}
\end{IEEEeqnarray}
We lower-bound the fourth term on the right-hand side of \eqref{eq:IYhat} as
\begin{IEEEeqnarray}{rCl}
\lefteqn{\frac{(b-1)(b-2)}{2}p_3 \log\frac{bp_3}{2p_2+(b-2)p_3}}~~~\nonumber\\
	& = & - \frac{(b-1)(b-2)}{2}p_3 \underbrace{\log\frac{2p_2+(b-2)p_3}{bp_3}}_{=\log\left(1+\frac{2(p_2-p_3)}{bp_3}\right) \le \frac{2(p_2-p_3)}{bp_3}\le \frac{2p_2}{bp_3}}\IEEEeqnarraynumspace \\
	& \ge & -\frac{(b-1)(b-2)}{2}\cdot \frac{2p_2}{b}\\
	& \ge & -bp_2. \label{eq:PPM2fourth}
\end{IEEEeqnarray}

Using \eqref{eq:exp} and \eqref{eq:exp2} we have the upper bound on $p_2$
\begin{IEEEeqnarray}{rCl}
	p_2 & = & \underbrace{(1-\e^{-\eta-c\ave})}_{\le \eta+c\ave} \underbrace{(1-\e^{-c\ave})}_{\le c\ave} \underbrace{\e^{-(b-2)c\ave}}_{\le 1} \\
	& \le & (\eta+c\ave) c\ave, \label{eq:p2up}
\end{IEEEeqnarray}
the lower bound on $p_2$
\begin{IEEEeqnarray}{rCl}
 	p_2 & = & \underbrace{(1-\e^{-\eta-c\ave)}}_{\ge 1-\e^{-\eta} \ge \eta-\frac{\eta^2}{2}}\underbrace{(1-\e^{-c\ave)}}_{\ge c\ave-\frac{c^2\ave^2}{2}} \underbrace{\e^{-(b-2)c\ave}}_{\ge 1-bc\ave}\\
	& \ge & \left(\eta-\frac{\eta^2}{2}\right)\left(c\ave-\frac{c^2\ave^2}{2}\right)(1-bc\ave), \label{eq:p2low}
\end{IEEEeqnarray}
and the upper bound on $p_3$
\begin{IEEEeqnarray}{rCl}
	p_3 & = &  \underbrace{\e^{-\eta-c\ave} }_{\le 1} {\underbrace{(1-\e^{-c\ave})}_{\le c\ave}}^2 \underbrace{\e^{-(b-3)c\ave}}_{\le 1}\\
	& \le & c^2\ave^2.\label{eq:p3up}
\end{IEEEeqnarray}

From \eqref{eq:IYhat},  \eqref{eq:PPM2third}, \eqref{eq:PPM2fourth}, \eqref{eq:p2up}, \eqref{eq:p2low}, and \eqref{eq:p3up} we obtain a lower bound on the additional mutual information that is gained by considering output frames with two detection positions:
\begin{IEEEeqnarray}{rCl}
	\lefteqn{I(\tilde{X};\hat{Y}) - I(\tilde{X};\tilde{Y})}~~~~~~\nonumber\\ & \ge & (b-1)p_2\log\frac{b}{2} - \frac{b^2}{2}p_3 - bp_2\\
	& \ge & (b-1)\left(\eta-\frac{\eta^2}{2}\right)\left(c\ave-\frac{c^2\ave^2}{2}\right)(1-bc\ave) \log\frac{b}{2} \nonumber\\
	& & {} - \frac{b^2}{2}c^2\ave^2 - b(\eta+c\ave) c\ave.
\end{IEEEeqnarray}
Dividing the above by $\eta$ and plugging in \eqref{eq:beta} we obtain a lower bound on the additional photon efficiency:
\begin{IEEEeqnarray}{rCl}
\lefteqn{\frac{I(\tilde{X};\hat{Y}) - I(\tilde{X};\tilde{Y})}{\eta}}~~~~~~\nonumber\\ & \ge & \left(1-\frac{\eta}{2}\right)(b-1) \left(c\ave - \frac{c^2 \ave^2}{2}\right) (1-c\eta) \log\frac{b}{2}\nonumber\\
& & {} - \frac{c^2}{2}\eta - c(\eta+c\ave).
 \label{eq:horror}
\end{IEEEeqnarray}
Adding the right-hand side of \eqref{eq:horror} to the right-hand side of \eqref{eq:prp1} yields \eqref{eq:prp2}.
\end{IEEEproof}


\section{Proof of the Upper Bounds
  \eqref{eq:upper1} and \eqref{eq:upper}}\label{sec:converse} 

\begin{proposition}\label{prp:upper}
Assume that $\ave$ is small enough so that
\begin{subequations}\label{eq:upper_con_all}
\begin{IEEEeqnarray}{rCl}
	\ave & < & \e^{-1} \label{eq:upper_cona}\\
	\ave\log\frac{1}{\ave} & < & \frac{\e^{-\frac{1}{1-2\e^{-1}}}}{1+c} \label{eq:upper_conb}\\
	\ave\left(\log\frac{1}{\ave}\right)^4 & < & \frac{144}{(1+c)^2}, \label{eq:upper_conc}
\end{IEEEeqnarray}
\end{subequations}
then
\begin{IEEEeqnarray}{rCl}
		C_\textnormal{PE}(\ave,c) & \le &  \log\frac{1}{\ave} - \log\log\frac{1}{\ave} - \log(1+c) + 2 + \log 13\nonumber\\
	& & {} + \frac{1}{\ave}\left( \log\frac{1}{1-(1+c)\ave} - (1+c)\ave\right) \nonumber\\
	& & {} + \frac{c^2}{2}\ave\log\log\frac{1}{\ave} + (1+c) \log\frac{1}{\displaystyle 1-\frac{1}{\log\frac{1}{\ave}}}. \IEEEeqnarraynumspace \label{eq:prp3}
	\end{IEEEeqnarray}
\end{proposition}

\begin{IEEEproof}[Proof of \eqref{eq:upper1} using Proposition~\ref{prp:upper}]
The last three summands on the right-hand side of~\eqref{eq:prp3} are all of the form $o(1)$. Setting $c=0$ in \eqref{eq:prp3} we have
\begin{IEEEeqnarray}{rCl}
	C_\textnormal{PE}(\ave,0) & \le & \log\frac{1}{\ave} - \log\log\frac{1}{\ave} + 2 + \log 13 + o(1)\IEEEeqnarraynumspace\\
	& = & \log\frac{1}{\ave} - \log\log\frac{1}{\ave} + O(1).
\end{IEEEeqnarray}
\end{IEEEproof}

\begin{IEEEproof}[Proof of \eqref{eq:upper} using Proposition~\ref{prp:upper}]
From \eqref{eq:prp3} we have
\begin{IEEEeqnarray}{rCl}
	\lefteqn{\varlimsup_{\ave\downarrow 0} \left\{ C_\textnormal{PE}(\ave,c) - \log\frac{1}{\ave} + \log\log\frac{1}{\ave}\right\}}~~~~~~~~\nonumber\\
	& \le & -\log(1+c) + 2 + \log 13.
\end{IEEEeqnarray}
Hence
\begin{IEEEeqnarray}{rCl}
	\lefteqn{\varlimsup_{c\to\infty} \varlimsup_{\ave\downarrow 0} \frac{\displaystyle C_\textnormal{PE}(\ave,c) - \log\frac{1}{\ave} + \log\log\frac{1}{\ave}}{\log c}}~~~~~~~~\nonumber\\
	& \le & \lim_{c\to\infty} \frac{-\log(1+c)+2+\log 13}{\log c}\\
	& = & -1.
\end{IEEEeqnarray}
\end{IEEEproof}

\begin{IEEEproof}[Proof of Proposition \ref{prp:upper}]
Like in
\cite{lapidothshapirovenkatesanwang11}, we use the duality
bound~\cite{lapidothmoser03_3} which states that, for any
distribution~$R(\cdot)$ on the output, the channel capacity satisfies 
\begin{equation}\label{eq:duality}
	C \le \sup \E{ D\bigl(W(\cdot|X) \| R(\cdot) \bigr)},
\end{equation}
where  the supremum is taken over all allowed input
distributions. We choose $R(\cdot)$ to be the following distribution:
\begin{equation}\label{eq:R}
  R(y) = \begin{cases} 1-(1+c)\ave, & y=0\\ \displaystyle (1+c) \ave \left( 1- \frac{1}{\log\frac{1}{\ave}}\right)\left(\frac{1}{\log\frac{1}{\ave}}\right)^{y-1}, &
    y \ge 1.\end{cases} 
\end{equation}

For any $x\ge 0$ we have
\begin{IEEEeqnarray}{rCl}
	\lefteqn{D\bigl( \bigl.W(\cdot|x)\bigr\| R(\cdot) \bigr)}\nonumber\\ & = & \sum_{y=0}^\infty \Poisson_{x+c\ave} (y) \cdot \log\frac{1}{R(y)} - H(Y|X=x)\\
	& = & \e^{-(x+c\ave)}\log\frac{1}{1-(1+c)\ave} \nonumber\\
	& &  {}- \sum_{y=1}^\infty \Poisson_{x+c\ave}(y) \nonumber\\
	& &  ~~~~~~~~{}\cdot \log \left\{ (1+c) \ave \left( 1- \frac{1}{\log\frac{1}{\ave}}\right)\left(\frac{1}{\log\frac{1}{\ave}}\right)^{y-1}\right\}\nonumber\\
	& & {} - H(Y|X=x)\\
	& = & \e^{-(x+c\ave)}\log\frac{1}{1-(1+c)\ave} + \underbrace{\sum_{y=1}^\infty \Poisson_{x+c\ave} (y) y}_{=\E{Y|X=x}=x+c\ave}  \log\log\frac{1}{\ave}\nonumber\\
	& & {} +  \underbrace{\left(\sum_{y=1}^\infty \Poisson_{x+c\ave} (y)\right)}_{=1-\e^{-(x+c\ave)}} \nonumber\\
	& & ~~~~~{}\cdot \left(\log\frac{1}{(1+c)\ave\log\frac{1}{\ave}} + \log\frac{1}{\displaystyle 1-\frac{1}{\log\frac{1}{\ave}}}\right)\nonumber\\
	& &  {} - H(Y|X=x)\\
	& = & \e^{-(x+c\ave)}\log\frac{1}{1-(1+c)\ave} + (x+c\ave)\log\log\frac{1}{\ave}\nonumber \\
	& & {}+(1-\e^{-(x+c\ave)})\nonumber\\
	& & ~~~~~{}\cdot\left(\log\frac{1}{(1+c)\ave\log\frac{1}{\ave}} + \log\frac{1}{\displaystyle 1-\frac{1}{\log\frac{1}{\ave}}}\right) \nonumber\\
	& & {}- H(Y|X=x). \label{eq:lower2_104}
\end{IEEEeqnarray}
We can lower-bound $H(Y|X=x)$ as
\begin{IEEEeqnarray}{rCl}
	H(Y|X=x) & \ge & \Hb (\e^{-(x+c\ave)})\\
	& = & \e^{-(x+c\ave)}(x+c\ave) \nonumber\\
	& & {} + (1-\e^{-(x+c\ave)})\log\frac{1}{1-\e^{-(x+c\ave)}}.\IEEEeqnarraynumspace
\end{IEEEeqnarray}
\pagebreak
Using this and \eqref{eq:lower2_104}, we upper-bound $D\bigl( \bigl.W(\cdot|x)\bigr\| R(\cdot) \bigr)$ as
\begin{IEEEeqnarray}{rCl}
	\lefteqn{D\bigl( \bigl.W(\cdot|x)\bigr\| R(\cdot) \bigr)}\nonumber\\
	 & \le & \e^{-(x+c\ave)}\log\frac{1}{1-(1+c)\ave} + (x+c\ave)\log\log\frac{1}{\ave}\nonumber \\
	& & {}+(1-\e^{-(x+c\ave)})\nonumber\\	
	& & ~~~~~{} \cdot \left(\log\frac{1}{(1+c)\ave\log\frac{1}{\ave}} + \log\frac{1}{\displaystyle 1-\frac{1}{\log\frac{1}{\ave}}}\right) \nonumber\\
	& & {} - \e^{-(x+c\ave)}(x+c\ave) - (1-\e^{-(x+c\ave)})\log\frac{1}{1-\e^{-(x+c\ave)}} \nonumber\\
	& & \, \\
	& = & \e^{-(x+c\ave)}\log\frac{1}{1-(1+c)\ave} + (x+c\ave)\log\log\frac{1}{\ave}\nonumber \\
	& & {} + \underbrace{(1-\e^{-(x+c\ave)})}_{\le x+c\ave} \underbrace{\log\frac{1}{\displaystyle 1-\frac{1}{\log\frac{1}{\ave}}}}_{\ge 0} - \e^{-(x+c\ave)} \underbrace{(x+c\ave)}_{\ge c\ave}\nonumber\\
	& & {} + (1-\e^{-(x+c\ave)}) \log\frac{1-\e^{-(x+c\ave)}}{(1+c)\ave\log\frac{1}{\ave}}\\
	& \le & \underbrace{\e^{-(x+c\ave)}}_{\le 1} \underbrace{\left( \log\frac{1}{1-(1+c)\ave} - c\ave\right)}_{\ge (1+c)\ave-c\ave \ge 0}\nonumber\\
	& & {} + (x+c\ave) \log\log\frac{1}{\ave} + (x+c\ave) \log\frac{1}{\displaystyle 1-\frac{1}{\log\frac{1}{\ave}}}\nonumber\\
	& & {} +\underbrace{(1-\e^{-(x+c\ave)})}_{=(1-\e^{-c\ave}) + (\e^{-c\ave} - \e^{-(x+c\ave)})} \log\frac{1-\e^{-(x+c\ave)}}{(1+c)\ave\log\frac{1}{\ave}}\\
	& \le & \left( \log\frac{1}{1-(1+c)\ave} - c\ave\right)\nonumber\\
	& & {} + (x+c\ave) \log\log\frac{1}{\ave} + (x+c\ave) \log\frac{1}{\displaystyle 1-\frac{1}{\log\frac{1}{\ave}}}\nonumber\\
	& & {} + (\e^{-c\ave} - \e^{-(x+c\ave)})\log\frac{1-\e^{-(x+c\ave)}}{(1+c)\ave\log\frac{1}{\ave}} \nonumber \\
	& & {} + (1-\e^{-c\ave}) \log\frac{1-\e^{-(x+c\ave)}}{(1+c)\ave\log\frac{1}{\ave}}\\
	& = & \ave + \left( \log\frac{1}{1-(1+c)\ave} - (1+c)\ave\right)\nonumber\\
	& & {} + (x+c\ave) \log\log\frac{1}{\ave} + (x+c\ave) \log\frac{1}{\displaystyle 1-\frac{1}{\log\frac{1}{\ave}}}\nonumber\\
	& & {} + \e^{-c\ave}(1 - \e^{-x})\log\frac{1-\e^{-(x+c\ave)}}{(1+c)\ave\log\frac{1}{\ave}} \nonumber \\
	& & {} + \underbrace{(1-\e^{-c\ave})}_{\ge c\ave - \frac{c^2}{2}\ave^2} \underbrace{\log\frac{c}{(1+c)\log\frac{1}{\ave}}}_{\le -\log\log\frac{1}{\ave}}\nonumber\\
	& & {} + \underbrace{(1-\e^{-c\ave})}_{\le c\ave}\underbrace{\log \frac{1-\e^{-(x+c\ave)}}{c\ave}}_{\le \log\frac{x+c\ave}{c\ave} \le \frac{x}{c\ave}}\label{eq:118}\\
	& \le & \ave + \left( \log\frac{1}{1-(1+c)\ave} - (1+c)\ave\right)\nonumber\\
	& & {} + (x+c\ave) \log\log\frac{1}{\ave} + (x+c\ave) \log\frac{1}{\displaystyle 1-\frac{1}{\log\frac{1}{\ave}}}\nonumber\\
	& & {} + \e^{-c\ave}(1 - \e^{-x})\log\frac{1-\e^{-(x+c\ave)}}{(1+c)\ave\log\frac{1}{\ave}} \nonumber \\
	& & {} - \left(c\ave - \frac{c^2}{2}\ave^2\right) \log\log\frac{1}{\ave} + x. \label{eq:long}
\end{IEEEeqnarray}
Using \eqref{eq:duality} and \eqref{eq:long} together with constraint \eqref{eq:ave} we have
\begin{IEEEeqnarray}{rCl}
	\lefteqn{C(\ave,c)}~~\nonumber\\
	 & \le & \sup \E{D\bigl( W(\cdot|X) \| R(\cdot) \bigr)}\\
	& \le & \ave + \left( \log\frac{1}{1-(1+c)\ave} - (1+c)\ave\right)\nonumber\\
	& & {} + (1+c)\ave \log\log\frac{1}{\ave} + (1+c)\ave \log\frac{1}{\displaystyle 1-\frac{1}{\log\frac{1}{\ave}}}\nonumber\\
	& & {} - \left(c\ave - \frac{c^2}{2}\ave^2\right) \log\log\frac{1}{\ave} + \ave\nonumber\\
	& & {} + \e^{-c\ave} \sup \E{(1 - \e^{-X})\log\frac{1-\e^{-(X+c\ave)}}{(1+c)\ave\log\frac{1}{\ave}}}\\
	& = & \ave\log\log\frac{1}{\ave} + 2\ave \nonumber\\
	& & {} + \left( \log\frac{1}{1-(1+c)\ave} - (1+c)\ave\right) \nonumber\\
	& & {} + \frac{c^2}{2}\ave^2\log\log\frac{1}{\ave} + (1+c)\ave\log\frac{1}{\displaystyle 1-\frac{1}{\log\frac{1}{\ave}}}\nonumber\\
	& & {} + \e^{-c\ave} \sup \E{(1 - \e^{-X})\log\frac{1-\e^{-(X+c\ave)}}{(1+c)\ave\log\frac{1}{\ave}}}. \IEEEeqnarraynumspace
\end{IEEEeqnarray}
Hence the maximum achievable photon efficiency satisfies
\begin{IEEEeqnarray}{rCl}
	\lefteqn{C_\textnormal{PE}(\ave,c)}~~~\nonumber\\
	& = & \frac{C(\ave,c)}{\ave}\\	
	& \le & \log\log\frac{1}{\ave} + 2 + \frac{1}{\ave}\left( \log\frac{1}{1-(1+c)\ave} - (1+c)\ave\right) \nonumber\\
	& & {} + \frac{c^2}{2}\ave\log\log\frac{1}{\ave} + (1+c) \log\frac{1}{\displaystyle 1-\frac{1}{\log\frac{1}{\ave}}}\nonumber\\
	& & {} + \frac{\e^{-c\ave}}{\ave}\sup \E{(1 - \e^{-X})\log\frac{1-\e^{-(X+c\ave)}}{(1+c)\ave\log\frac{1}{\ave}}}.\label{eq:CPE23} \IEEEeqnarraynumspace
\end{IEEEeqnarray}

We next upper-bound the expectation on the right-hand side of \eqref{eq:CPE23} as follows
\begin{IEEEeqnarray}{rCl}
	\lefteqn{\E{(1 - \e^{-X})\log\frac{1-\e^{-(X+c\ave)}}{(1+c)\ave\log\frac{1}{\ave} } }}~~~~~~\nonumber\\
	& = & \E{X \cdot \frac{1-\e^{-X}}{X}\log\frac{1-\e^{-(X+c\ave)}}{(1+c)\ave\log\frac{1}{\ave}}}\\
	& \le & \E{X \cdot \frac{1-\e^{-X}}{X}\log\frac{X+c\ave}{(1+c)\ave\log\frac{1}{\ave}}}\\
	& \le & \ave \cdot \sup \phi(x) \label{eq:E121}
\end{IEEEeqnarray}
where
\begin{equation}\label{eq:127}
	\phi(x) \triangleq \frac{1-\e^{-x}}{x}\log\frac{x+c\ave}{(1+c)\ave\log\frac{1}{\ave}},\quad x\ge 0.
\end{equation}

We claim that, if $\ave$ is small enough to satisfy all of the conditions in \eqref{eq:upper_con_all},
then
\begin{equation}\label{eq:12log}
	\sup\phi(x) = \phi(x^*) \quad \textnormal{for some }x^* \le \frac{12}{\log\frac{1}{\ave}}.
\end{equation}

To show this, we compute the derivative of $\phi(x)$ (with respect to $x$) as
\begin{equation} \label{eq:derivative}
	\frac{\d \phi(x)}{\d x} = \frac{(1+x)\e^{-x}-1}{x^2}\log\frac{x+c\ave}{(1+c)\ave\log\frac{1}{\ave}} + \frac{1-\e^{-x}}{x(x+c\ave)}
\end{equation}
which exists and is continuous for all $x>0$. Further note that it is negative for large enough $x$, and is positive for small enough $x$. Hence $x^*$ must be achieved at a point where
\begin{equation}
	\frac{\d \phi(x)}{\d x} = 0.
\end{equation}
To prove \eqref{eq:12log}, it thus suffices to show that, under the assumptions \eqref{eq:upper_con_all},
\begin{equation}
	\frac{\d \phi(x)}{\d x} <0,\quad x >\frac{12}{\log\frac{1}{\ave}}.
\end{equation}
We do this separately for two cases.

\begin{case} Consider
\begin{equation}
	\frac{12}{\log\frac{1}{\ave}} < x < 1.
\end{equation}
Note that this case need not be considered if $\ave\ge \e^{-12}$.
\end{case}
In this case the logarithmic term in the derivative is positive:
\begin{IEEEeqnarray}{rCl}
	\log\frac{x+c\ave}{(1+c)\ave\log\frac{1}{\ave}} & \ge & \log\frac{x}{(1+c)\ave\log\frac{1}{\ave}}\\
	& > & \log\frac{12}{(1+c)\ave\left(\log\frac{1}{\ave}\right)^2}\\
	& > & 0, \label{eq:case1_1}
\end{IEEEeqnarray}
where the last step follows because \eqref{eq:upper_cona} and \eqref{eq:upper_conc} together imply
\begin{equation}\label{eq:upper_cond}
	\ave\left(\log\frac{1}{\ave}\right)^2 < \frac{12\e^{-\frac{1}{2}}}{1+c}.
\end{equation}
Next, using Taylor's theorem with Cauchy remainder, we have
\begin{IEEEeqnarray}{rCl}
	\frac{(1+x)\e^{-x} - 1}{x^2} & = & -\frac{1}{2} + \frac{1}{3}x -\underbrace{\frac{(3-x')\e^{-x'}}{24} x^2 }_{\ge 0} \label{eq:cauchy1} \IEEEeqnarraynumspace \\
	& \le & -\frac{1}{2} + \frac{1}{3} \underbrace{x}_{\le 1} \label{eq:cauchy2}\\
	& \le & -\frac{1}{6}, \label{eq:case1_2}
\end{IEEEeqnarray}
where $x' \in [0,x]$. The underbrace in \eqref{eq:cauchy1} follows because $x'\le x< 1$.
We also have
\begin{equation}\label{eq:case1_3}
	\frac{1-\e^{-x}}{x(x+c\ave)} \le \frac{x}{x(x+c\ave)} \le \frac{1}{x}.
\end{equation}
Plugging \eqref{eq:case1_1}, \eqref{eq:case1_2} and \eqref{eq:case1_3} into \eqref{eq:derivative} we have
\begin{IEEEeqnarray}{rCl}
	\frac{\d \phi(x)}{\d x} & \le & -\frac{1}{6} \log\frac{x+c\ave}{(1+c)\ave\log\frac{1}{\ave}} + \frac{1}{x}\\
	& \le & -\frac{1}{6} \underbrace{\log x}_{\ge \log 12 - \log\log\frac{1}{\ave}} + \frac{1}{6}\log(1+c) - \frac{1}{6} \log\frac{1}{\ave}\nonumber\\
	& & {} + \frac{1}{6}\log\log\frac{1}{\ave} + \underbrace{\frac{1}{x}}_{\le \frac{1}{12}\log\frac{1}{\ave}}\\
	& \le & \frac{1}{6}\log\frac{1+c}{12} + \frac{1}{3}\log\log\frac{1}{\ave} - \frac{1}{12}\log\frac{1}{\ave}\\
	& = & \frac{1}{12} \log\left( \frac{(1+c)^2}{144} \ave \left(\log\frac{1}{\ave}\right)^4 \right)\\
	& < & 0,
\end{IEEEeqnarray}
where the last step follows from \eqref{eq:upper_conc}. We have now shown
\begin{equation}\label{eq:case1}
	\frac{\d \phi(x)}{\d x} < 0,\quad \frac{12}{\log\frac{1}{\ave}} < x < 1.
\end{equation}

\begin{case} Consider
\begin{equation}
	x\ge 1.
\end{equation}
\end{case}
In this case the logarithmic term in the derivative is still positive:
\begin{equation}\label{eq:case2_1}
	\log\frac{x+c\ave}{(1+c)\ave\log\frac{1}{\ave}} \ge \log\frac{1}{(1+c)\ave\log\frac{1}{\ave}} > 0,
\end{equation}
where the last step follows from \eqref{eq:upper_conb}. We bound the other terms as
\begin{equation}\label{eq:case2_2}
\frac{(1+x)\e^{-x} - 1}{x^2} \le - \frac{1 - 2 \e^{-1}}{x^2}
\end{equation}
which is negative, and
\begin{equation}\label{eq:case2_3}
	\frac{1-\e^{-x}}{x(x+c\ave)} \le \frac{1}{x^2}.
\end{equation}
Using \eqref{eq:case2_1}, \eqref{eq:case2_2} and \eqref{eq:case2_3} together with \eqref{eq:derivative} we have
\begin{IEEEeqnarray}{rCl}
	\frac{\d \phi(x)}{\d x} & \le & - \frac{1-2\e^{-1}}{x^2}\left(\underbrace{\log(x+c\ave)}_{\ge0} + \log\frac{1}{(1+c)\ave\log\frac{1}{\ave}}\right) \nonumber\\
	& & {} + \frac{1}{x^2}\\
	& \le & -\frac{1-2\e^{-1}}{x^2} \log\frac{1}{(1+c)\ave\log\frac{1}{\ave}} \nonumber\\
	& & {} + \frac{1-2\e^{-1}}{x^2} \cdot\frac{1}{1-2\e^{-1}}\\
	& = & -\frac{1-2\e^{-1}}{x^2} \log\frac{\e^{-\frac{1}{1-2\e^{-1}}}}{(1+c)\ave\log\frac{1}{\ave}} \\
	& < & 0,
\end{IEEEeqnarray}
where the last step follows from \eqref{eq:upper_conb}. Hence we have shown
\begin{equation} \label{eq:case2}
\frac{\d \phi(x)}{\d x} < 0,\quad x\ge 1.
\end{equation}

Combining \eqref{eq:case1} and \eqref{eq:case2} proves \eqref{eq:12log} for all $\ave$ satisfying~\eqref{eq:upper_con_all}.

We now proceed to upper-bound $\sup \phi(x)$:
\begin{IEEEeqnarray}{rCl}
	\sup \phi(x) & = & \phi(x^*)\\
	& = & \underbrace{\frac{1-\e^{-x^*}}{x^*}}_{\le 1} \log\frac{x^* + c\ave}{(1+c)\ave\log\frac{1}{\ave}}\\
	& \le & \max \left\{ 0, \log\frac{x^*+c\ave}{(1+c)\ave\log\frac{1}{\ave}}\right\}\\
	& \le & \max \left\{ 0, \log\frac{\displaystyle \frac{12}{\log\frac{1}{\ave}}+c\ave}{(1+c)\ave\log\frac{1}{\ave}} \right\}. \label{eq:upper147}
\end{IEEEeqnarray}
Now note that, due to \eqref{eq:upper_conb}, we have
\begin{equation}
	c\ave \le \frac{1}{\log\frac{1}{\ave}}.
\end{equation}
We can now continue the chain of inequalities \eqref{eq:upper147} as
\begin{IEEEeqnarray}{rCl}
	\sup \phi(x) & \le & \max\left\{0, \log\frac{13}{(1+c)\ave\left(\log\frac{1}{\ave}\right)^2} \right\} \\
	& = & \log\frac{13}{(1+c)\ave\left(\log\frac{1}{\ave}\right)^2} \label{eq:upper_45}\\
	& = & \log \frac{1}{\ave} - 2\log\log\frac{1}{\ave} - \log(1+c) + \log 13. \label{eq:upper_140} \IEEEeqnarraynumspace
\end{IEEEeqnarray}
Here \eqref{eq:upper_45} follows because conditions \eqref{eq:upper_cona} and \eqref{eq:upper_conc} imply \eqref{eq:upper_cond}.

Combining \eqref{eq:CPE23}, \eqref{eq:E121} and \eqref{eq:upper_140}, and noting $\e^{-c\ave}\le 1$ (together with the fact that the right-hand side of \eqref{eq:upper_140} is positive), we obtain
\begin{IEEEeqnarray}{rCl}
	\lefteqn{C_\textnormal{PE} (\ave,c)}~~~\nonumber\\
	 & \le & \log\log\frac{1}{\ave} + 2 + \frac{1}{\ave}\left( \log\frac{1}{1-(1+c)\ave} - (1+c)\ave\right) \nonumber\\
	& & {} + \frac{c^2}{2}\ave\log\log\frac{1}{\ave} + (1+c) \log\frac{1}{\displaystyle 1-\frac{1}{\log\frac{1}{\ave}}}\nonumber\\
	& & {} +\log \frac{1}{\ave} - 2\log\log\frac{1}{\ave} - \log(1+c) + \log 13\\
	& = & \log\frac{1}{\ave} - \log\log\frac{1}{\ave} - \log(1+c) + 2 + \log 13\nonumber\\
	& & {} + \frac{1}{\ave}\left( \log\frac{1}{1-(1+c)\ave} - (1+c)\ave\right) \nonumber\\
	& & {} + \frac{c^2}{2}\ave\log\log\frac{1}{\ave} + (1+c) \log\frac{1}{\displaystyle 1-\frac{1}{\log\frac{1}{\ave}}}.
\end{IEEEeqnarray}
\end{IEEEproof}

\section{Numerical Comparison and Concluding Remarks}\label{sec:conclusion}
\begin{figure}[tbp]
        \centering
        \psfrag{E}{\footnotesize $\mathcal{E}$}
        \psfrag{Upper bound}[Bl][Bl]{\scriptsize Upper bound}
        \psfrag{Approximation}[Bl][Bl]{\scriptsize Approximation \eqref{eq:log1c}}
        \psfrag{Simple PPM}[Bl][Bl]{\scriptsize Simple PPM}
	\psfrag{Soft-decision PPM}[Bl][Bl]{\scriptsize Soft-decision PPM}
        \psfrag{On-off keying}[Bl][Bl]{\scriptsize On-off keying}
	\psfrag{Photon Efficiency (nats/photon)}[Bl][Bl]{\scriptsize Photon Efficiency (nats/photon)}
	\subfigure[Case $c=0.1$. Note that the lowest three curves almost overlap.]{\includegraphics[width=0.4\textwidth]{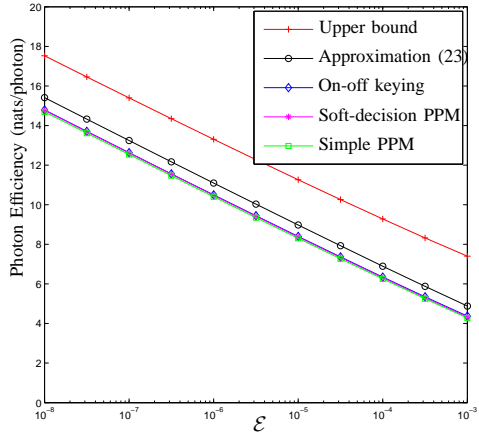}
       }
    \subfigure[Case $c=1$. The on-off keying and soft-decision PPM curves overlap.]{\includegraphics[width=0.4\textwidth]{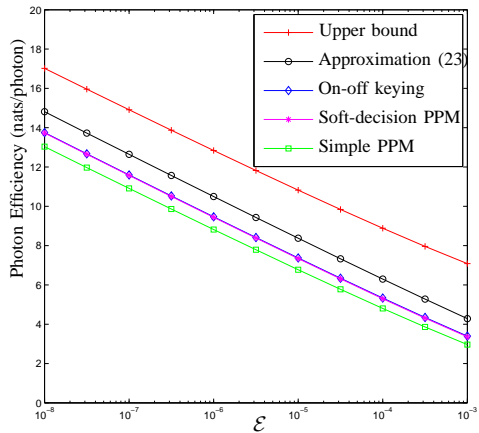}
}	
\subfigure[Case $c=10$.]{\includegraphics[width=0.4\textwidth]{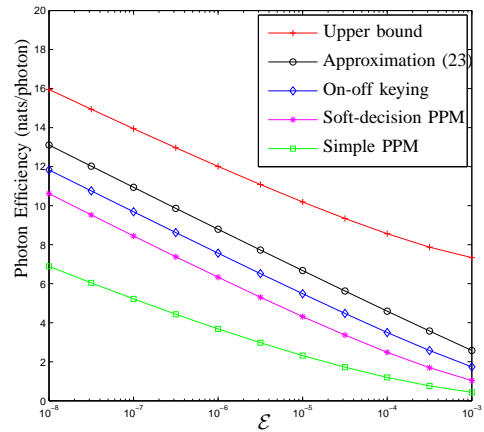}}
        \caption{Comparing the approximation~\eqref{eq:log1c} 
          to nonasymptotic upper
          bound, and to lower bounds for
          simple PPM, soft-decision PPM, and on-off keying for three cases: $c=0.1$, $c=1$, and $c=10$. } \label{fig:1}  
    \end{figure}

We numerically compare the approximation \eqref{eq:log1c} with nonasymptotic upper and lower bounds on photon efficiency. Specifically, the plotted upper bound is based on \eqref{eq:CPE23} and \eqref{eq:E121}, and is given by
\begin{IEEEeqnarray}{rCl}
	\lefteqn{C_\textnormal{PE} (\ave,c)}~~~\nonumber\\ 	
	& \le & \log\log\frac{1}{\ave} + 2 + \frac{1}{\ave}\left( \log\frac{1}{1-(1+c)\ave} - (1+c)\ave\right) \nonumber\\
	& & {} + \frac{c^2}{2}\ave\log\log\frac{1}{\ave} + (1+c) \log\frac{1}{\displaystyle 1-\frac{1}{\log\frac{1}{\ave}}}\nonumber\\
	& & {} + \e^{-c\ave} \sup \phi(x),
\end{IEEEeqnarray}
where $\phi(\cdot)$ is given in \eqref{eq:127} and $\sup \phi(x)$ is computed numerically.
The on-off-keying lower bound is obtained by computing the mutual information of the channel with ``on'' signal equaling $\eta$, and with the receiver ignoring multiple detected photons. This is given by
\begin{IEEEeqnarray}{rCl}
	\lefteqn{C_\textnormal{PE-OOK} (\ave,c)}~~\nonumber\\
	& \ge & \frac{1}{\ave} \left(H_\textnormal{b} (q) - \frac{\ave}{\eta} \cdot H_\textnormal{b} (\e^{-\eta-c\ave}) - \left(1-\frac{\ave}{\eta}\right) H_\textnormal{b} (\e^{-c\ave})\right),\nonumber\\
	& & \,  \label{eq:OOKplot}
\end{IEEEeqnarray}
where $H_\textnormal{b}(\cdot)$ denotes the binary entropy function
\begin{equation}
	H_\textnormal{b}(a) = a\log\frac{1}{a} + (1-a)\log\frac{1}{1-a},\quad a\in[0,1],
\end{equation}
and where
\begin{equation}
	q \triangleq \frac{\ave}{\eta}\e^{-\eta-c\ave} + \left(1-\frac{\ave}{\eta}\right) \e^{-c\ave}.
\end{equation}
The simple-PPM lower bound is computed using \eqref{eq:super} and \eqref{eq:PPM_exact}:
\begin{IEEEeqnarray}{rCl}
	C_\textnormal{PE-PPM} (\ave,c) & \ge & \frac{1}{\eta} \left( p_0 \log\frac{b p_0}{p_0+(b-1)p_1} \right.\nonumber\\
	& & ~~~~{}+ \left.(b-1) p_1\log
  \frac{bp_1}{p_0+(b-1)p_1} \right).\label{eq:PPM1plot} \IEEEeqnarraynumspace
\end{IEEEeqnarray} 
The soft-decision-PPM lower bound is computed using \eqref{eq:IYhat} and is given by
\begin{IEEEeqnarray}{rCl}
	\lefteqn{C_\textnormal{PE-PPM(SD)} (\ave,c)}~~\nonumber\\
	 & \ge & \frac{1}{\eta} \left(p_0\log\frac{bp_0}{p_0+(b-1)p_1}\right.\nonumber\\
	& & ~~~~{} + (b-1)p_1\log\frac{bp_1}{p_0+(b-1)p_1}\nonumber\\
	& & ~~~~{} + (b-1)p_2 \log\frac{bp_2}{2p_2+(b-2)p_3}\nonumber\\
	& & ~~~~{} + \left.\frac{(b-1)(b-2)}{2}p_3 \log\frac{bp_3}{2p_2+(b-2)p_3}\right). \label{eq:PPM2plot} \IEEEeqnarraynumspace
\end{IEEEeqnarray}
In all these lower bounds, the parameters $b$ and $\eta$ are chosen as in \eqref{eq:b} and \eqref{eq:beta}, and $p_0$, $p_1$, $p_2$, $p_3$, and $p_4$ are given in \eqref{eq:defp0}, \eqref{eq:defp1}, \eqref{eq:defp2}, \eqref{eq:defp3}, and \eqref{eq:defp4}, respectively.
We plot these bounds for $c=0.1$, $c=1$, and $c=10$ in Figure~\ref{fig:1}.

The figure shows the following.
\begin{itemize}
	\item The approximation \eqref{eq:log1c} is reasonably accurate for small enough $\ave$.
	\item The on-off-keying and the soft-decision-PPM bounds are consistently close to the approximation \eqref{eq:log1c}.
	\item As $c$ increases, the simple-PPM bound deviates significantly from the other lower bounds, and hence also from the actual value of $C_\textnormal{PE}(\ave,c)$.
\end{itemize}


The capacity bounds as well as the asymptotic results in this paper provide insights to how technical restrictions and device or channel imperfections influence the communication rates in optical channels in the wideband regime. Our results and techniques can also be applied to secret communication and key distribution over optical channels \cite{kochmanwangwornell13}. It would be interesting to see if these are further extendable, e.g., to multiple-mode optical channels and optical networks.

As we have demonstrated, PPM is nearly optimal on the Poisson channel in the high-photon-efficiency regime. When $c$ is small, the simple-PPM super channel has high erasure probability but low ``error'' (by which we mean the receiver detects a single pulse at a position that is different from the transmitted signal) probability. In this case, Reed-Solomon codes can perform rather close to the theoretical limit. However, when $c>1$, Reed-Solomon codes can no longer achieve any positive rates on this channel. Nevertheless, we believe that, for $c>1$, PPM still has its advantages over on-off keying in terms of code design. This is because the optimal input distribution for (both simple and soft-decision) PPM is uniform, whereas the optimal input distribution for on-off keying for this channel is highly skewed. The uniformity of PPM inputs allows the usage of more structured codes, in particular \emph{linear codes}. One possible direction, for instance, is to employ the idea of \emph{multilevel codes} \cite{imaihirakawa77,wachsmannfischerhuber99} on this channel.

\appendix

In this appendix we provide a proof for the converse part of \eqref{eq:lsvw} that is much simpler than the one given in \cite{lapidothshapirovenkatesanwang11}. Using the same arguments as in \cite{lapidothshapirovenkatesanwang11}, we know that $C(\ave,c)$ is monotonically increasing in $c$. Hence it suffices to show
\begin{equation}
	\varlimsup_{\ave\downarrow 0} \frac{C(\ave,0)}{\ave\log\frac{1}{\ave}} \le 1.
\end{equation}
This can be verified as follows:
\begin{IEEEeqnarray}{rCl}
	C(\ave,0) & = & \max_{\E{X}\le \ave} I(X;Y)\\
	& \le & \max_{\E{X}\le \ave} H(Y)\\
	& = & \max_{\E{Y}\le \ave} H(Y)\\
	& = & (1+\ave)\log(1+\ave) - \ave\log\ave, \label{eq:boseeinstein}
\end{IEEEeqnarray}
which yields the desired asymptotic bound. Here, \eqref{eq:boseeinstein} follows from the well-known fact that, for a nonnegative discrete random variable under a first-moment constraint, maximum entropy is achieved by the geometric distribution, and is given by the right-hand side of \eqref{eq:boseeinstein}. Note that the right-hand side of \eqref{eq:boseeinstein} is exactly the capacity of the pure-loss bosonic channel~\cite{giovannettiguha04}.

\section*{Acknowledgments}
The authors thank Hye Won Chung, Yuval Kochman, Amos Lapidoth, Arya Mazumdar,
and Lizhong Zheng for helpful discussions, and the anonymous reviewers for their careful reading of the manuscript.

\bibliographystyle{IEEEtran}           
\bibliography{/Volumes/Data/wang/Library/texmf/tex/bibtex/header_short,/Volumes/Data/wang/Library/texmf/tex/bibtex/bibliofile}

\end{document}